\documentclass[preprint,12pt]{elsarticle}

\usepackage{amssymb}
\usepackage{amsmath}
\usepackage{parskip}
\usepackage{multirow}

\usepackage{pdfpages}
\usepackage{graphicx}
\usepackage{graphics}
\usepackage {framed}
\usepackage {amsxtra}
\usepackage {enumerate}
\usepackage{commath}
\usepackage[margin=1 in]{geometry}
\usepackage{float}
\usepackage[caption = false]{subfig}
\usepackage{bm}
\usepackage[utf8]{inputenc}
\usepackage[english]{babel}
\title{A Simple Illustration of Interleaved Learning \\ using Kalman Filter for Linear Least Squares}

\journal{arXiv}

\begin{document}

\author{Majnu John, Yihren Wu}

\begin{abstract}
Interleaved learning in machine learning algorithms is a biologically inspired training method with promising results. In this short note, we illustrate the interleaving mechanism via a simple statistical and optimization framework based on Kalman Filter for Linear Least Squares.
\end{abstract}


\maketitle

\section{Introduction}

Interleaved learning (IL) is a type of biological learning phenomenon observed in brain regions such as neocortex, and has inspired machine learning algorithms. IL is one of the mechanisms expounded by Complementary Learning Systems Theory (McClelland, McNaughton and O'Reilly, 1995; Marr, 1971) on how successful learners such as human beings mitigate effects of `catastrophic interference' while learning.  Recent illustrations of IL using neural networks include Saxena, Shobe and McNaughton, 2022, who exhibited that if the new information is similar to a subset of old items, then deep neural networks can learn the new information rapidly and with the same level of accuracy by interleaving the old items in the subset. A similar insight was presented in McClelland, McNaughton and Lampinen, 2020, where it was shown that for artificial neural networks, information consistent with prior knowledge can sometimes be integrated very quickly. Another recent paper (Ban and Xie, 2021) formulated interleaved machine learning as a multi-level optimization problem, and developed an efficient differentiable algorithm to solve the interleaving learning problem with application to neural architecture search. A closely related biological concept is interleaved replay which also has been empirically validated in the literature (Gepperth and Karaoguz, 2016; Kemker and Kanan, 2018).  Over the past couple of decades, ideas inspired by biological IL have been utilized in a wide array of online learning methods as well, especially to prevent catastrophic forgetting. See, for example Wang \textit{et. al.}, 2015, a comprehensive and recently updated survey on continual, lifelong learning.

All the applications and illustrations of IL to machine learning so far have used complex models such as neural networks. In this short paper, we aim to present a simple illustration of IL by adapting a framework from traditional optimization and statistics literature, namely, the Kalman-Filter (KF) approach for linear least squares (LLS). Understanding IL in this relatively straightforward framework may help in future with proving theoretical convergence properties, and then with hopefully extending to similar results for more complex models and gradient descent type algorithms. To the best of our knowledge, an illustration of IL based on Kalman filter for linear least squares, has not yet appeared in the literature (e.g., no mention of such an approach in the comprehensive survey by Wang \textit{et al.}, 2015).

 \section{Kalman Filter for Linear Least Squares (KF4LLS)}

 To better understand the concepts and notation later on, we briefly review KF4LLS by closely following the exposition provided in section 3.2 in Bertsekas and Tsitsiklis, 1996. Note that although KF is widely considered as an estimation method associated with dynamical systems, the one employed for linear least squares is a specialized case where the states of the underlying dynamical system stays constant (Bertsekas, 1996).

 Consider fitting a linear model to the set of input-output pairs $(\mathbf{y}_{1}, \mathbf{X}_{1}), \ldots, (\mathbf{y}_{m}, \mathbf{X}_{m})$. Here $\mathbf{y}_{i} \in \mathbb{R}^{n_{i}}$, $\mathbf{X}_{i}$ is an $n_{i} \times q$ matrix and each $(\mathbf{y}_{i}, \mathbf{X}_{i})$ is a given data block. Model fitting corresponds to minimizing the following quadratic cost function \[ \mathcal{C}(\mathbf{r}) = \sum_{i=1}^{m}||\mathbf{y}_{i} - \mathbf{X}_{i}\mathbf{r}||^{2}, \;\; \mathrm{for}\;\; \mathbf{r} \in \mathbb{R}^{q}. \]

 KF4LLS is an incremental version of Gauss-Newton method, which cycles through the data blocks. Specifically, the solution given by KF4LLS to the above minimization program is $\boldsymbol \psi_{m}$ which can be obtained recursively by the algorithm \[ \boldsymbol \psi_{i} = \boldsymbol \psi_{i-1} + \mathbf{H}_{i}^{-1}\mathbf{X}_{i}^{t}(\mathbf{y}_{i} - \mathbf{X}_{i}\boldsymbol \psi_{i-1}), \;\; \mathbf{H}_{i} = \mathbf{H}_{i-1} + \mathbf{X}_{i}^{t}\mathbf{X}_{i}, \;\; i = 1, \ldots, m, \] where $\boldsymbol \psi_{0}$ is an arbitrary vector and $\mathbf{H}_{0} = \mathbf{0}$. We assume that $\mathbf{X}_{1}^{t}\mathbf{X}_{1}$ is positive definite and that makes all $\mathbf{H}_{i}$'s (except $\mathbf{H}_{0}$) positive definite as well. KF4LSS has been well-studied in the literature (see, for example, papers citing Bertsekas, 1996). A derivation of the algorithm is given in section 3.2 in Bertsekas and Tsitsiklis, 1996 and convergence analysis is presented in Bertsekas, 1996.

  \section{Interleaving KF4LLS}

  Consider data blocks from two different `populations' \[(\mathbf{b}_{1}, \mathbf{U}_{1}), \ldots, (\mathbf{b}_{m}, \mathbf{U}_{m}) \;\; \mathrm{and} \;\; (\mathbf{f}_{1}, \mathbf{V}_{1}), \ldots, (\mathbf{f}_{m}, \mathbf{V}_{m}).\] To fix ideas, it may help to think in terms of an example provided in McClelland, McNaughton and O'Reilly, 1995, where the two populations are birds and fish. In our notation, we may think of columns of $\mathbf{U}_{i}$'s and $\mathbf{V}_{i}$'s as features related to birds and fish, respectively, and similarly $\mathbf{b}_{i}$'s and $\mathbf{f}_{i}$'s the corresponding target variables. In previous papers that mentioned this example, the target variables were class variables but in this paper, for convenience and simplicity, we focus on continuously distributed target variables.

  We also consider a third population which is a `mixture' of the first two populations in terms of data characteristics. In our running example, the third population will be penguins. In terms of features, we think of penguins as an admixture of birds and fish - they have wings like a bird and they can swim like a fish$!$ We denote the data blocks from the penguin population as \[ (\mathbf{p}_{1}, \mathbf{Z}_{1}), \ldots, (\mathbf{p}_{m}, \mathbf{Z}_{m}). \] For all populations, we assume the relationship between the corresponding target variables and features data to be linear models: \[ \mathbf{b}_{i} = \mathbf{U}_{i}\mathbf{r}_{b} + \varepsilon_{b}, \; \;  \mathbf{f}_{i} = \mathbf{V}_{i}\mathbf{r}_{f} + \varepsilon_{f}, \; \;  \mathbf{p}_{i} = \mathbf{Z}_{i}\mathbf{r}_{p} + \varepsilon_{p}, \; i = 1, \ldots, m, \] where $\varepsilon_{b}, \varepsilon_{f}$ and $\varepsilon_{p}$ are \textit{i.i.d.} mean zero error variables, and for convenience, we assume $m = 2k$ (\textit{i.e.}, an even number). We are interested in knowing whether we can train a model by interleaving data blocks from birds and fish, and then use the trained model to predict target variables related to penguins (\textit{i.e.} the $\mathbf{p}_{i}$'s) using features data from penguins (\textit{i.e.} the $\mathbf{Z}_{i}$'s). That is, the goal is to train a model via interleaving, using only data from birds and fish, but then test the model using only features and target variables from penguins.

  In this short note, we assume \begin{equation} \label{mixeq1} \mathbf{Z}_{i} = \alpha \mathbf{U}_{i} + (1 - \alpha)  \mathbf{V}_{i} \end{equation} and \begin{equation} \label{mixeq2} \mathbf{r}_{p} = \alpha \mathbf{r}_{b} + (1-\alpha)\mathbf{r}_{f}, \;\mathrm{for\;some\;} \alpha \in [0,1]. \end{equation} In words, each feature matrix $\mathbf{Z}_{i}$ of penguins is a weighted average of $\mathbf{U}_{i}$ and $\mathbf{V}_{i}$, the corresponding feature matrices of birds and fish. Similarly, the weight-parameters in the model for penguins, $\mathbf{r}_{p}$, connecting the features to the target variable is a weighted average of the weight-parameters in the models for birds and fish. If instead of (\ref{mixeq1}), we assumed the distribution of $\mathbf{Z}_{i}$'s to be a mixture of distributions of $\mathbf{U}_{i}$'s and $\mathbf{V}_{i}$'s we observed similar results as the ones presented later in this short note, but to focus the presentation we will just work with the assumption made in (\ref{mixeq1}). In this case, our interleaved algorithm is as follows.

  \noindent \textbf{\underline{Interleaved KF4LLS algorithm}:}\\

  \noindent \textbf{\textit{Step 0(a)}:} Center all data blocks, including the target variables, individually by subtracting the corresponding column means. Thus, in the following step, all $\mathbf{b}_{i}$'s, $\mathbf{f}_{i}$'s, and all columns of $\mathbf{U}_{i}$'s and $\mathbf{V}_{i}$'s are mean-zero vectors.

  \noindent \textbf{\textit{Step 0(b)}:} Set $\mathbf{H}_{0}^{(\alpha)} = \mathbf{0}$ and \[\mathbf{U}_{i}^{(\alpha)} = \sqrt{\alpha}\,\mathbf{U}_{i}, \; \mathbf{b}_{i}^{(\alpha)} = \sqrt{\alpha}\,\mathbf{b}_{i};\; \;\;\mathbf{V}_{i}^{(\alpha)} = \sqrt{1-\alpha}\,\mathbf{V}_{i}, \; \mathbf{f}_{i}^{(\alpha)} = \sqrt{1-\alpha}\,\mathbf{f}_{i}, \;i = 1, \ldots, m. \]

  \noindent \textbf{\textit{Step 1}:} $\mathbf{H}_{1}^{(\alpha)} =  \mathbf{H}_{0}^{(\alpha)} + (\mathbf{U}_{1}^{(\alpha)})^{t}(\mathbf{U}_{1}^{(\alpha)}); \; \boldsymbol \psi_{1} = \boldsymbol \psi_{0} + (\mathbf{H}_{1}^{(\alpha)})^{-1}(\mathbf{U}_{1}^{(\alpha)})^{t}(\mathbf{b}_{1}^{(\alpha)} - \mathbf{U}_{1}^{(\alpha)}\boldsymbol \psi_{0}). $\\
  \noindent \textbf{\textit{Step 2}:} $\mathbf{H}_{2}^{(\alpha)} =  \mathbf{H}_{1}^{(\alpha)} + (\mathbf{V}_{1}^{(\alpha)})^{t}(\mathbf{V}_{1}^{(\alpha)}); \; \boldsymbol \psi_{2} = \boldsymbol \psi_{1} + (\mathbf{H}_{2}^{(\alpha)})^{-1}(\mathbf{V}_{1}^{(\alpha)})^{t}(\mathbf{f}_{1}^{(\alpha)} - \mathbf{V}_{1}^{(\alpha)}\boldsymbol \psi_{1}). $\\
  \noindent \textbf{\textit{Step 3}:} $\mathbf{H}_{3}^{(\alpha)} =  \mathbf{H}_{2}^{(\alpha)} + (\mathbf{U}_{2}^{(\alpha)})^{t}(\mathbf{U}_{2}^{(\alpha)}); \; \boldsymbol \psi_{3} = \boldsymbol \psi_{2} + (\mathbf{H}_{3}^{(\alpha)})^{-1}(\mathbf{U}_{2}^{(\alpha)})^{t}(\mathbf{b}_{2}^{(\alpha)} - \mathbf{U}_{2}^{(\alpha)}\boldsymbol \psi_{2}). $\\
  \noindent \textbf{\textit{Step 4}:} $\mathbf{H}_{4}^{(\alpha)} =  \mathbf{H}_{3}^{(\alpha)} + (\mathbf{V}_{2}^{(\alpha)})^{t}(\mathbf{V}_{2}^{(\alpha)}); \; \boldsymbol \psi_{4} = \boldsymbol \psi_{3} + (\mathbf{H}_{4}^{(\alpha)})^{-1}(\mathbf{V}_{2}^{(\alpha)})^{t}(\mathbf{f}_{2}^{(\alpha)} - \mathbf{V}_{2}^{(\alpha)}\boldsymbol \psi_{3}). $\\
  ... etc. ... \\
  \noindent \textbf{\textit{Step (m-1)}:} $\mathbf{H}_{m-1}^{(\alpha)} =  \mathbf{H}_{m-2}^{(\alpha)} + (\mathbf{U}_{k}^{(\alpha)})^{t}(\mathbf{U}_{k}^{(\alpha)}); \; \boldsymbol \psi_{m-1} = \boldsymbol \psi_{m-2} + (\mathbf{H}_{m-1}^{(\alpha)})^{-1}(\mathbf{U}_{k}^{(\alpha)})^{t}(\mathbf{b}_{k}^{(\alpha)} - \mathbf{U}_{k}^{(\alpha)}\boldsymbol \psi_{m-2})$, where $k = m/2$.\\
  \noindent \textbf{\textit{Step m}:} $\mathbf{H}_{m}^{(\alpha)} =  \mathbf{H}_{m-1}^{(\alpha)} + (\mathbf{V}_{k}^{(\alpha)})^{t}(\mathbf{V}_{k}^{(\alpha)}); \; \boldsymbol \psi_{m} = \boldsymbol \psi_{m-1} + (\mathbf{H}_{m}^{(\alpha)})^{-1}(\mathbf{V}_{k}^{(\alpha)})^{t}(\mathbf{f}_{k}^{(\alpha)} - \mathbf{V}_{k}^{(\alpha)}\boldsymbol \psi_{m-1})$.\

  The algorithm alternates between using $(\mathbf{b}_{i}, \mathbf{U}_{i})$'s (i.e. birds data blocks) in odd-numbered steps and $(\mathbf{f}_{i}, \mathbf{V}_{i})$'s (i.e. fish data blocks) in even-numbered steps making it a proper interleaved training approach. Note that the algorithm is an oracle algorithm because it can be implemented only if the mixing coefficient $\alpha$ is known. Typically this is possible only for simulated synthetic data. Thus, the above algorithm in its current form serves only for illustrating IL and not for any practical applications. For real data, if assumption (\ref{mixeq1}) truly holds then $\alpha$ can be estimated. It is more likely that for real data there will be a separate mixing coefficient for each column of $\mathbf{Z}_{i}$; such separate coefficients can also be estimated, for example, using a grid search on the unit interval.

  \noindent \textbf{Illustration with synthetic data}

  We illustrate the algorithm on synthetic data generated as follows. We set $\alpha = 0.25$, $n_{i} = n = 100$, $q = 6$ and $m = 6$ (i.e. $k$ = 3).  Performance of the algorithm was assessed by calculating the bias and the mean-squared error (MSE) based on the estimates $\boldsymbol \psi_{i}$'s after each step of the algorithm. Average bias and MSE over 5000 simulation-iterations were plotted (see Figure 1). Elements in $\mathbf{r}_{b}$ and $\mathbf{r}_{f}$ were generated separately from $Uniform(-5, 5)$ distribution, and then fixed for all simulation iterations. For each simulation-iteration, each row of $\mathbf{U}_{i}$ was generated independently from a multivariate normal - $N(\boldsymbol \mu_{1}, I_{6 \times 6})$, and similarly each row of $\mathbf{V}_{i}$ was generated from $N(\boldsymbol \mu_{2}, I_{6 \times 6})$ where $\boldsymbol \mu_{1} = [\mu_{1}, \ldots, \mu_{1}]^{t}$ and $\boldsymbol \mu_{2} = [\mu_{2}, \ldots, \mu_{2}]^{t}$. Here $\mu_{1}$, $\mu_{2}$ were generated separately from a $Uniform(-10, 10)$ distribution, and then fixed for all the simulation-iterations. Bias plotted below was averaged across all simulation-iterations, but within each simulation-iteration it was also averaged across elements of the parameter vector. Codes used for this example with detailed comments are posted in the following GitHub page (https://github.com/mjohn5/InterleavedKF4LLS/)

    \begin{figure}[H]
    \begin{center}
    \hspace*{-0.75cm}
    \includegraphics[height = 3.5in, width = 7in, angle = 0]{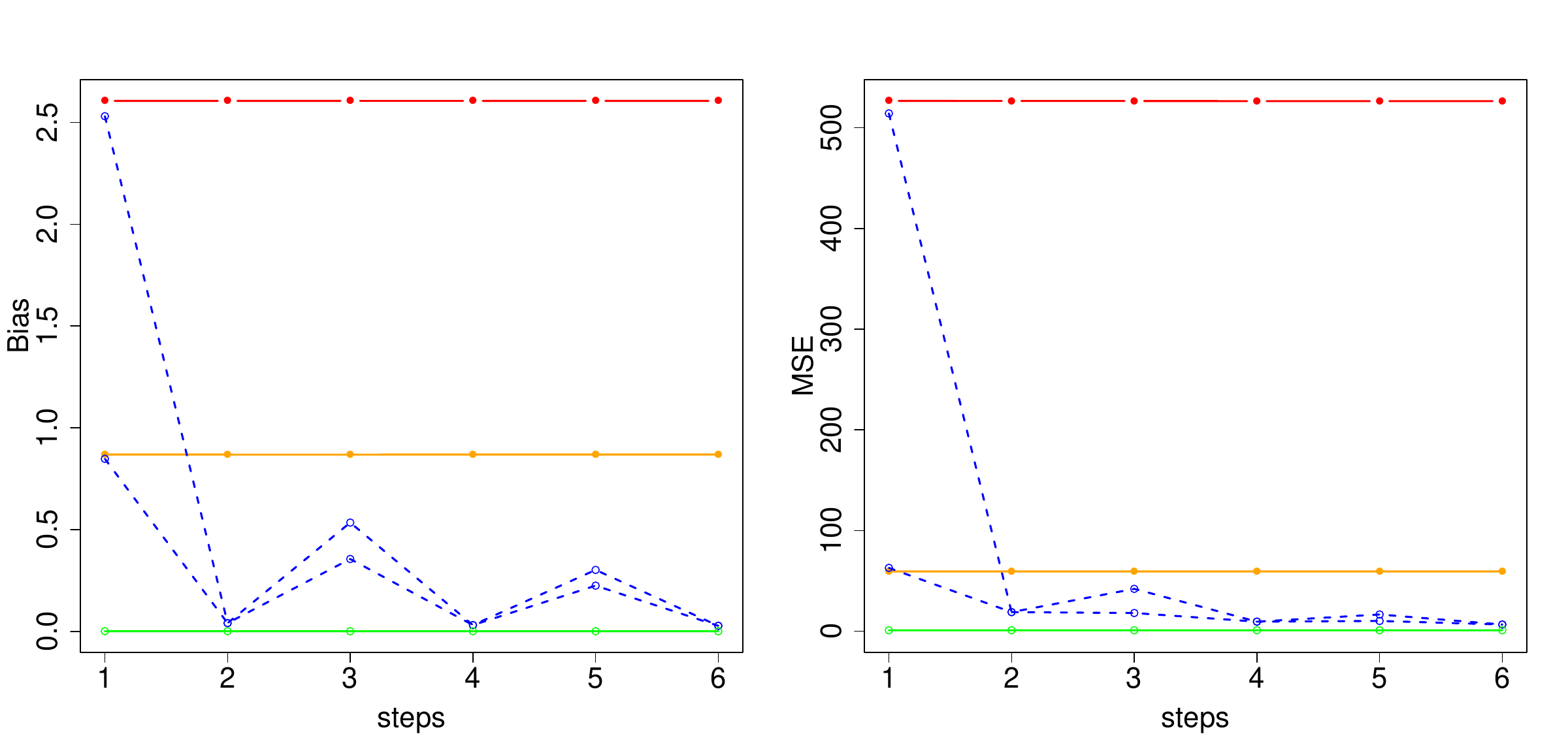}
    \caption{Bias and MSE, averaged across 5000 simulation-iterations, of interleaved KF4LLS algorithm applied on synthetic data. In all scenarios, MSE was calculated as the prediction error when the trained models where applied on `penguin' test data. Red, orange and green lines correspond to training based on birds, fish and penguin data blocks, respectively, without interleaving. The blue lines correspond to training based on interleaving algorithm, either starting with a bird data block or with a fish data block.}
    \end{center}
    \end{figure}

  The red line in Figure 1 corresponds to the scenario where only the $(\mathbf{b}_{i}, \mathbf{U}_{i})$ blocks were used for training, and the orange line corresponds to the scenario where only the $(\mathbf{f}_{i}, \mathbf{V}_{i})$ blocks were used for training. Since all the testing was done on $(\mathbf{p}_{i}, \mathbf{Z}_{i})$ blocks, it is not surprising to see that the scenarios corresponding to the red and orange lines show substantial bias and MSE for all steps. The green line corresponds to the scenario at the other extreme where training and testing were both done on the penguin data (i.e. $(\mathbf{p}_{i}, \mathbf{Z}_{i})$). Again, it is not surprising to see that the bias and MSE for this scenario is very close to zero. Blue lines correspond to the scenario with Interleaved KF4LLS algorithm used for training, and as in all other cases, testing done on `penguin' blocks. There are two blue lines in each panel, one starting with $(\mathbf{b}_{1}, \mathbf{U}_{1})$ and the other starting with $(\mathbf{f}_{1}, \mathbf{V}_{1})$; in both cases the algorithm alternates between the `birds' and `fish' data blocks. It is easy to see from the figure that, similar to the biological interleaved learning phenomenon, interleaving the training in this simple least squares setting leads to almost nil bias and MSE. The reduction in bias and MSE achieved by the Interleaved KF4LLS algorithm in a few steps is almost the same as that achieved by the algorithm that is trained exclusively with `penguin' data. With this synthetic data example, it is also observed that the Interleaved KF4LLS algorithm achieves almost zero bias in just two steps, a phenomenon that has some theoretical justification (see below).

  \noindent \textbf{Some theoretical justification}

  Let $\mathcal{F}_{2j}$ denote the `history of the algorithm' up to and including the $2j^{th}$ step, $j = 1, \ldots, k$. That is, $\mathcal{F}_{2j}$ is the sigma-field generated by $\mathbf{b}_{1}, \mathbf{U}_{1}, \ldots, \mathbf{b}_{j}, \mathbf{U}_{j};\; \mathbf{f}_{1}, \mathbf{V}_{1}, \ldots, \mathbf{f}_{j}, \mathbf{V}_{j}$. Then the following lemmas show that even with two steps the estimator obtained by the algorithm (i.e. $\boldsymbol \psi_{2}$) is a good approximation to the unknown parameter-vector that we are trying to estimate, namely, $\mathbf{r}_{p}$. Thus, the following theory closely mirrors the result that we saw with synthetic data above.

  \noindent \textbf{Lemma-1.} \begin{equation} \label{psi2l1} \boldsymbol \psi_{2} = (\mathbf{H}_{2}^{(\alpha)})^{-1}[(\mathbf{U}_{1}^{(\alpha)})\mathbf{b}_{1} + (\mathbf{V}_{1}^{(\alpha)})\mathbf{f}_{1}]. \end{equation} Hence, \begin{equation} \label{epsi2l1} \mathbb{E}(\boldsymbol \psi_{2}/ \mathcal{F}_{2}) = \left[\alpha (\mathbf{U}_{1}^{t}\mathbf{U}_{1}) + (1-\alpha)(\mathbf{V}_{1}^{t}\mathbf{V}_{1})\right]^{-1}\left[\alpha (\mathbf{U}_{1}^{t}\mathbf{U}_{1})\mathbf{r}_{b} + (1-\alpha)(\mathbf{V}_{1}^{t}\mathbf{V}_{1})\mathbf{r}_{f}\right]. \end{equation}

  \noindent \textbf{Proof of Lemma-1:} Adding \[ \mathbf{H}_{1}^{(\alpha)}\boldsymbol \psi_{1} = \mathbf{H}_{1}^{(\alpha)}\boldsymbol \psi_{0} + (\mathbf{U}_{1}^{(\alpha)})^{t}\mathbf{b}_{1}^{(\alpha)} - (\mathbf{U}_{1}^{(\alpha)})^{t}\mathbf{U}_{1}^{(\alpha)}\boldsymbol \psi_{0} \] and \[ \mathbf{H}_{2}^{(\alpha)}\boldsymbol \psi_{2} = \mathbf{H}_{2}^{(\alpha)}\boldsymbol \psi_{1} + (\mathbf{V}_{1}^{(\alpha)})^{t}\mathbf{f}_{1}^{(\alpha)} - (\mathbf{V}_{1}^{(\alpha)})^{t}\mathbf{V}_{1}^{(\alpha)}\boldsymbol \psi_{1} \] and cancelling terms, we get eq. (\ref{psi2l1}). Eq. (\ref{epsi2l1}) follows from eq. (\ref{psi2l1}) since $\mathbb{E}(\mathbf{b}_{1}/ \mathcal{F}_{2}) = \mathbf{U}_{1}\mathbf{r}_{b}$ and  $\mathbb{E}(\mathbf{f}_{1}/ \mathcal{F}_{2}) = \mathbf{V}_{1}\mathbf{r}_{f}$.

  Also, as a side remark, the symmetry in the result above explains why it is irrelevant whether we start with $(\mathbf{b}_{1}, \mathbf{U}_{1})$ or with $(\mathbf{f}_{1}, \mathbf{V}_{1})$ as seen in the synthetic data example. The following lemma states that up to a first order approximation based on Taylor series expansion, $\boldsymbol \psi_{2}$ calculated in step-2 of the Interleaving KF4LLS algorithm is an unbiased estimator of $\mathbf{r}_{p}$, if the columns of $\mathbf{U}_{1}$ (and similarly columns of $\mathbf{V}_{1}$) are (respectively) pairwise uncorrelated and with constant standard deviation.

  \noindent \textbf{Lemma-2.} If \begin{equation} \label{e1l2} n^{-1}\mathbb{E}(\mathbf{U}_{1}^{t}\mathbf{U}_{1}) = n^{-1}\mathbb{E}(\mathbf{V}_{1}^{t}\mathbf{V}_{1}) = \sigma I_{n \times n}, \end{equation} then up to a first order approximation \begin{equation} \label{e2l2} \mathbb{E}(\boldsymbol \psi_{2}) \approx \alpha \mathbf{r}_{b} + (1-\alpha)\mathbf{r}_{f} = \mathbf{r}_{p}. \end{equation}

   \noindent \textbf{Proof of Lemma-2:} It is well-known that using a first-order Taylor series approximation, the expectation of a ratio is approximately the ratio of the expectation. Taking expectations in eq. (\ref{epsi2l1}), applying the above-mentioned fact and using eq. (\ref{e1l2}) we get eq. (\ref{e2l2}).

 \section{Conclusions, Brief Discussion and Future Directions}

   Interleaved learning is a learning technique observed in human brain areas such as neocortex which helps with long-term retention and in general better learning. Inspired by this biological phenomenon, machine learning algorithms have tried to incorporate interleaving while training models, especially complex neural network models. In this short note, we presented a simple statistical framework based on linear least squares to better understand computational interleaving learning.

   Our assumption in eq. (\ref{mixeq1}), we think, makes intuitive sense. However at first glance, it may seem that our assumption in eq. (\ref{mixeq2}) on the weight parameters is a bit artificial or unrealistic from a real data perspective, especially since we use the same mixing coefficient $\alpha$ as in eq. (\ref{mixeq1}). We would like to point out that interleaving based machine learning algorithms seen in the literature (see for example Ban and Xie, 2021) make the assumption that weight parameters are same or similar across the `learners'. For example, in Ban and Xie, 2021, the authors go even to the extend of incorporating a penalty term in the optimization program to ensure that the weight parameters do not vary across the learners, which is based on the implicit assumption that the true weight parameters in the underlying unknown model are the same across learners. Translating into our setting, we have two `learners' corresponding to the data blocks from birds and fish. If we assume that the weight parameters for these two learners to be the same (that is, $\mathbf{r}_{b} = \mathbf{r}_{f}$), then our assumption in eq. (\ref{mixeq2}) forces \begin{equation}\label{parameq} \mathbf{r}_{p} = \mathbf{r}_{b} = \mathbf{r}_{f}. \end{equation} The theory that we presented showcasing the approximate unbiasedness of $\boldsymbol \psi_{2}$ follows with eq. (\ref{parameq}) as well which is a special case of eq. (\ref{mixeq2}).
   
   An important comment related to eq. (\ref{mixeq1}) is that implicitly we assumed the dimensions of $\mathbf{U}_{i}$'s and $\mathbf{V}_{i}$'s to be the same, making our setting a bit restrictive. Also, although not explicitly stated, we think that in our set-up the corresponding columns of $\mathbf{U}_{i}$'s and $\mathbf{V}_{i}$'s have to be both numerical from a real data analysis perspective. If one or both of them are categorical, interpreting $\mathbf{Z}_{i}$ based on eq. (\ref{mixeq1}) may often be problematic in a practical setting. In our synthetic data example, we had the corresponding columns in $\mathbf{U}_{i}$'s and $\mathbf{V}_{i}$'s not only numerical but also from the same distribution with only the mean values different, which may make our example a bit limited. However, we note that some similarity between the corresponding columns in $\mathbf{U}_{i}$'s and $\mathbf{V}_{i}$'s is necessary for interleaving to be effective. It has been mentioned in previous literature (Saxena, Shobe and McNaughton, 2022; McClelland, McNaughton and Lampinen, 2020) that the biological brain interleaves only old items with substantial representational similarity to new items.

   Our simple framework based on linear least squares can probably be extended to logistic regression models or any generalized linear models and support vector machines as well, which we intend to pursue as future work. A framework like the one presented in this short note will also help with better understanding the convergence properties of interleaving algorithms. Future work will include stating and proving such theoretical properties as well.

\section*{Acknowledgements} We thank Professor Bruce McNaughton for inspiring us on this line of work. We also thank Rajat Saxena and Bruce McNaughton for sharing their draft review paper related to forward transfer in continual learning.

\section*{References}

  Ban, H., Xie, P. (2021). Interleaving Learning, with Application to Neural Architecture Search. \textit{arXiv preprint.} arXiv:2103.07018.

  Bertsekas, D.P., Tsitsiklis, J.N. (1996). Neuro-Dynamic Programming. \textit{Athena Scientific, Bellmont, MA.} $4^{th}$ printing.

  Bertsekas, D.P. (1996). Incremental least squares methods and the extended Kalman filter. \textit{SIAM J. Optim.} 6(3), 807–822.

  Gepperth, A., Karaoguz, C. (2016). A bio-inspired incremental learning architecture for applied perceptual
problems. \textit{Cognit. Comput.} 8, 924–934.

  Kemker, R., Kanan, C. (2017). FearNet: Brain-Inspired Model for Incremental Learning. \textit{ICLR poster 2018.} arXiv: 1711.10563.

  Marr, D. (1971). Simple memory: A theory for archicortex. \textit{Philos. Trans. R. Soc. Lond. B Biol. Sci.} 262, 23–81.

  McClelland, J.L., McNaughton, B.L., Lampinen, A.K. (2020). Integration of new information in memory: New insights from a complementary learning systems perspective. \textit{Philos. Trans. R. Soc. Lond. B Biol. Sci.} 375, 20190637.

  McClelland, J.L., McNaughton, B.L., O'Reilly, R.C. (1995). Why there are complementary learning systems in the hippocampus and neocortex: Insights from the successes and failures of connectionist models of learning and memory. \textit{Psychol. Rev.} 102, 419–457.

  Saxena, R., Shobe, J.L., McNaughton, B.L. (2022) Learning in deep neural networks and brains with similarity-weighted interleaved learning. \textit{Proc Natl Acad Sci U S A.} 119(27): e2115229119. doi: 10.1073/pnas.2115229119.

  Wang, L., Zhang, X., Su, H., Zhu, J. (2023). A comprehensive survey of continual learning: Theory, method and application. \textit{arXiv preprint}. arXiv:2302.00487.

\section*{Affiliations}

  MJ: Departments of Mathematics and of Psychiatry, Hofstra University, Hempstead, NY.
  YW: Department of Mathematics, Hofstra University, Hempstead, NY.

\end{document}